# Coupled electronic and magnetic excitations in the cuprates and their role in the superconducting transition

F. Restrepo[1*], U. Chatterjee[2], G. D. Gu[3], H. Xu[1], D. K. Morr[1,4], and J. C. Campuzano[1*]

[1]Deptartment of Physics, University of Illinois at Chicago, Chicago, Illinois 60607, USA

[2]Department of Physics, University of Virginia, Charlottesville, Virginia 22904, USA

[3]Condensed Matter Physics and Materials Science Division, Brookhaven National Laboratory, Upton, New York 11973, USA

[4]James Franck Institute, University of Chicago, Chicago, Illinois 60637, USA

[*]Correspondence and requests for materials should be addressed to J.C.C. (email: jcc@uic.edu) or F.R. (email: frestr2@uic.edu)

**The formation of Cooper pairs, a bound state of two electrons of opposite spin and momenta by exchange of a phonon [1], is a defining feature of conventional superconductivity. In the cuprate high temperature superconductors, even though it has been established that the superconducting state also consists of Cooper pairs, the pairing mechanism remains intensely debated. Here we investigate superconducting pairing in the $Bi_2Sr_2CaCu_2O_{8+\delta}$ (Bi2212) cuprate by employing spectral functions obtained directly from angle-resolved photoemission (ARPES) experiments as input to the Bethe-Salpeter gap equation. Assuming that Cooper pairing is driven solely by spin fluctuations, we construct the single-loop spin-fluctuation-mediated pairing interaction, and use it to compute the eigenfunctions and eigenvalues of the Bethe-Salpeter equation in the particle-particle channel for multiple Bi2212 samples. The key point of our results is that, as the temperature is reduced, the leading eigenvalue increases upon approaching $T_c$, reaching a value of approximately 1 at the $T_c$ corresponding to each doping value, indicating a superconducting transition with $d_{x^2-y^2}$-wave eigenfunctions. This suggests that spin fluctuations can approximately account for $T_c$ and, consequently, mediate pairing in the cuprate high temperature superconductors.**

Superconductivity in the cuprates emerges following quantum melting of the antiferromagnetic Mott insulating state of the parent compound, either via electron or hole doping. It has been proposed that the spin fluctuations resulting from the melted Mott state act as the "glue" leading to Cooper pairing in high temperature superconductors [2, 3]. In support of this proposal, neutron scattering experiments on the $YBa_2Cu_3O_{6.95}$ cuprate superconductors demonstrate that the change in magnetic exchange energy between the superconducting and normal states can provide sufficient superconducting condensation energy [4]. Moreover, it has been shown that the interaction of electrons with spin fluctuations can account for a number of anomalies in charge, spin, and optical response measurements in the cuprates [3]. However, an outstanding question is whether the spin-fermion interaction is strong enough to yield the high transition temperatures observed in these materials.

While several studies have attempted to answer this important question by using a combination of theoretical approaches and various experimentally measured electronic and spin excitation spectra, the results have been inconclusive. Using the electron spectral functions from ARPES experiments in near-optimally doped Bi2212, Mishra et al. [5] solved a generalized gap equation, but did not find a superconducting transition for any temperature. In contrast, Dahm et al. [6] employed the spin excitation spectrum measured via inelastic neutron scattering experiments (INS) in $YBa_2Cu_3O_{6.6}$ to obtain a superconducting transition temperature of about 150 K. However, in this study, the electron spectral function was theoretically computed, and not taken from experimental ARPES data. In a subsequent theoretical study, Maier et al. [7] ascribed the failure of Mishra et al. [5] to find any superconducting transition to the fact that the experimental ARPES data were taken at a temperature of 140 K, well above the critical temperature of $T_C = 90$ K. Taken together, these studies have suggested that, in order to determine whether the spin-fermion interaction can be the source of the unconventional superconducting state in the cuprate superconductors, three requirements need to be satisfied: (i) experimental ARPES data need to be taken above, but near $T_c$, (ii) the effective superconducting pairing vertex needs to be computed from these ARPES data, and (iii) the solution of the gap equation, using the ARPES data and the derived pairing vertex as input, should reproduce the transition temperature of the material.

Here, we use ARPES data from three underdoped and optimally doped Bi2212 samples, taken at temperatures near and above $T_c$, to construct the spin-fluctuation-mediated pairing vertex and solve the generalized gap equation. This yields a superconducting order parameter with $d_{x^2-y^2}$ −wave symmetry, and a critical temperature for all samples which is close to the ones observed experimentally.

Following previous work [5-8], we employ the Bethe-Salpeter equation for a generalized order parameter $\Phi(\mathbf{k},\omega_n,T)$:

$$\lambda(T)\Phi(T) = \hat{O}(T)\Phi(T). \tag{1}$$

The transition to superconductivity takes place when the eigenvalue $\lambda$ reaches $\lambda(T_c) = 1$. The operator $\hat{O}(T)$ is given by (we solve the equation in Matsubara frequency space)

$$\hat{O}(T)\Phi(\mathbf{k},\omega_n,T) = -\frac{T}{N}\sum_{\mathbf{k}',\Omega_n} V(\mathbf{k}-\mathbf{k}',i\omega_n-i\Omega_n,T)G(\mathbf{k}',i\Omega_n,T)G(-\mathbf{k}',-i\Omega_n,T)\Phi(\mathbf{k}',\Omega_n,T), \tag{2}$$

where $G$ is the electron's Green's function and $V$ the effective electron-electron pairing interaction. Equations (1) and (2) are quite universal in that the specific mechanism leading to the emergence of superconductivity is reflected only in the form of $V$, as noted previously by Mishra et al. [5].

The required quantities for the Bethe-Salpeter equation can be obtained, within certain approximations, from the electron spectral function $A(\mathbf{k},\omega)$ obtained from the ARPES signal $I(\mathbf{k},\omega) = I_0 f(\omega)A(\mathbf{k},\omega) + B(\omega)$, where $I_0$ is the matrix element, $f(\omega)$ the Fermi function, and $B(\omega)$ the inelastic background signal. To extract the spectral function, we first subtract the background signal, while simultaneously normalizing out the matrix element $I_0$ [9]. We then note that, although ARPES only measures the occupied states, the cuprates exhibit particle-hole mixing below [10] and above [11] $T_c$. Therefore, the symmetrized spectral function $A(\mathbf{k},\omega) = I(\mathbf{k},\omega) + I(-\mathbf{k}+2\mathbf{k}_F,-\omega)$ also yields the unoccupied part, for $\mathbf{k}$ values close to the Fermi momentum $\mathbf{k}_F$. The Green's functions in the Matsubara representation then follow from the relation

$$G(\mathbf{k}, \omega_n, T) = \int d\omega \, \frac{A(\mathbf{k},\omega,T)}{i\omega_n - \omega}. \tag{3}$$

We analyze ARPES data from three Bi2212 samples: two thin films, one underdoped with $T_c = 67$ K (UD67), the other optimally doped with $T_c = 80$ K (OP80), and a single crystal, optimally doped with $T_c = 91$ K (OP91). Representative data are shown in Fig. 1. Figure 1a shows the experimentally determined phase diagram for our samples [12], with the circles, diamonds, and squares indicating the spectra used in this work. In Fig.1b, we plot the spectral function in the Y-quadrant of the Brillouin zone, with the Fermi surface appearing as a continuous yellow line. The measured spectra for the UD67 sample along the high symmetry directions, highlighted by the coloured dots in panel b, are shown in Fig. 1c.

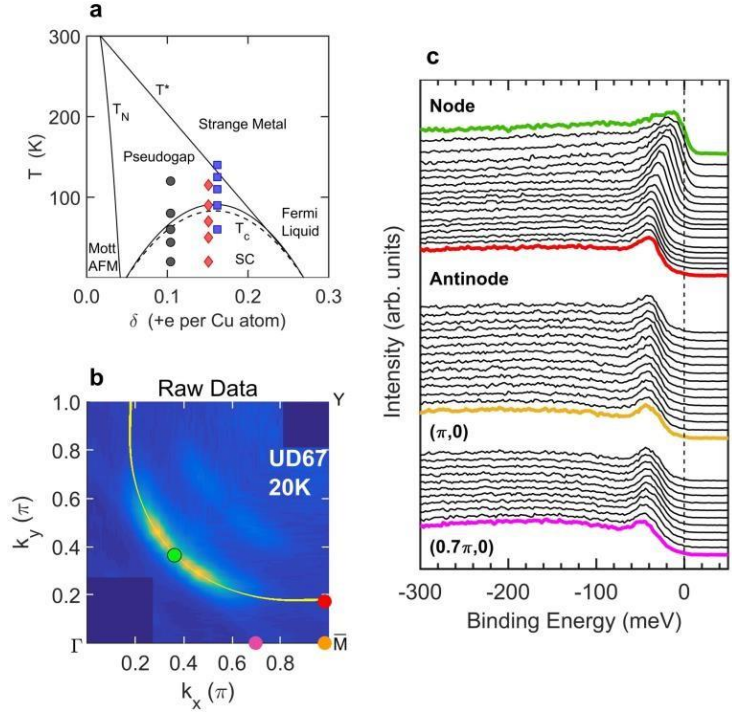

Figure 1: Extent and resolution of ARPES data employed in calculation of the pairing interaction $V$ and the Green's function $G$. (**a**) Phase diagram of Bi2212. Black circles correspond to the UD67 sample, red diamonds to OP80, and blue squares to OP91. There are two superconducting domes, one dashed for the thin films and one solid for the single crystal, with a higher optimal $T_c$. (**b**) Raw Fermi Surface (after interpolation and reflection about symmetry axes) for the UD67 sample at 20

K. The thin yellow line is a tight-binding fit to the Bi2212 dispersion [13] and is used here as a guide to the eye. (**c**) Energy Distribution Curves corresponding to panel (**b**) along the path indicated by the coloured dots. Throughout this article, we take $a = 1$ for the lattice constant.

For pairing induced by spin fluctuations, we consider the pairing vertex [14]

$$V(\mathbf{q}, i\omega_m) = \frac{3}{2} U_0^2 \chi(\mathbf{q}, i\omega_m), \tag{4}$$

where $\chi$ is the spin susceptibility and $U_0$ is an effective spin-fermion coupling energy. The spectral function allows us to calculate the (real frequency) spin fluctuation propagator in the Random Phase Approximation (RPA) [15]:

$$\chi(\mathbf{q}, \omega) = \frac{\chi_0(\mathbf{q}, \omega)}{1 - U_q \chi_0(\mathbf{q}, \omega)}, \tag{5}$$

where $\chi_0$ is the bare spin susceptibility,

$$\chi_0(\mathbf{q}, \Omega) = \frac{1}{N} \sum_{\mathbf{k}} \int_{-\infty}^{\infty} d\omega d\nu A(\mathbf{k}, \omega) A(\mathbf{k}+\mathbf{q}, \nu) \frac{f(\nu) - f(\omega)}{\Omega + \nu - \omega + i\delta}, \tag{6}$$

and the coupling $U_q$ is assumed to have a super-exchange momentum dependence [13, 16–18]:

$$U_q = -\frac{U_0}{2}(\cos q_x + \cos q_y) \tag{7}$$

(see Supplementary Information). Since $U_0$ is a high energy scale not provided by the RPA nor the low-energy formalism adopted here, its value was chosen so as to reproduce the salient features of the spin susceptibility measured in INS experiments. In particular, in the superconducting state, the cuprates exhibit a sharp resonance in their spin excitation spectra at the commensurate wave-vector $\mathbf{Q} = (\pi,\pi)$ [19, 20]. The energy of this resonance throughout the phase diagram of the cuprate YBa$_2$Cu$_3$O$_{6+\delta}$ has been extensively reported; however, due to technical limitations, such information is lacking for underdoped Bi2212. For this reason, we used the empirical relation [20–22]

$$E_R = 5.4 k_B T_c \qquad (8)$$

to fix the resonance energies for the three samples ($E_R^{67}$ =32 meV, $E_R^{80}$ =37 meV, and $E_R^{91}$ =41 meV). We then chose $U_0$ such that Im $\chi(\mathbf{Q},\Omega)$ exhibited resonance peaks at the energies given by equation (8) (for a different estimate of the coupling constant, see Supplementary Information). As the temperature increases, INS data suggest that the resonance broadens and decreases in intensity, but disperses weakly with temperature, even above $T_c$ [23–25]. This observation, which is consistent with the reported near-constancy of the ARPES antinodal gap in the temperature range considered here [26], justifies the use of equation (8), even for the underdoped Bi2212 compounds where no INS data are available. Near $T_c$, the values of $U_0$ extracted from this procedure are ~ 730 meV for UD67, ~ 650 meV for OP80 and ~ 580 meV for OP91 (see Supplementary Information for a discussion of $U_0$ and the contribution from the anomalous Green's function).

Figure 2 shows the calculated spin susceptibilities (Im$\chi$) for the two Bi2212 films (additional results for the single crystal can be found in Supplementary Information). (Here we include results well below $T_c$ to compare with the more extensive INS data in this regime.) The commensurate peak intensity decreases with increasing temperature and doping (Figs. 2a,c), as seen in INS experiments on YBa$_2$Cu$_3$O$_{6+x}$ [27]. Also, the resonance peak widths deep in the superconducting state are ~ 15 − 20 meV (blue curves in Figs. 2a,c), in agreement with INS data from optimally doped Bi2212 [19]. In addition to the resonance, our calculations also reproduce the ubiquitous

"upward branch" ($\omega > E_R$), observed in all cuprate families, but not the "lower branch", which is material-specific [6, 28] (Figs. 2b,d). The energy-momentum integral of Im$\chi(\mathbf{q},\omega)$ (multiplied by the matrix element $2\mu_B^2$ [13]) gives a total fluctuating moment of $\langle m^2 \rangle \sim 0.43\mu_B^2$ for the UD67 sample and $\langle m^2 \rangle \sim 0.39\mu_B^2$ for OP80, both varying weakly with temperature. Moreover, for the UD67 sample at 20 K, the total spectral weight taken up by the resonance (integrated over all momenta and from 0 to 70 meV) is about $0.16\mu_B^2$, whereas the remaining spectral weight (up to 200 meV), $\sim 0.27\mu_B^2$, is taken up by high-energy magnetic excitations. These values compare reasonably well with the corresponding figures for YBa$_2$Cu$_3$O$_{6.6}$ (with a similar $T_c$) [29] and suggest that the high-energy structure of the spin response must be included in a description of the magnetically mediated pairing interaction.

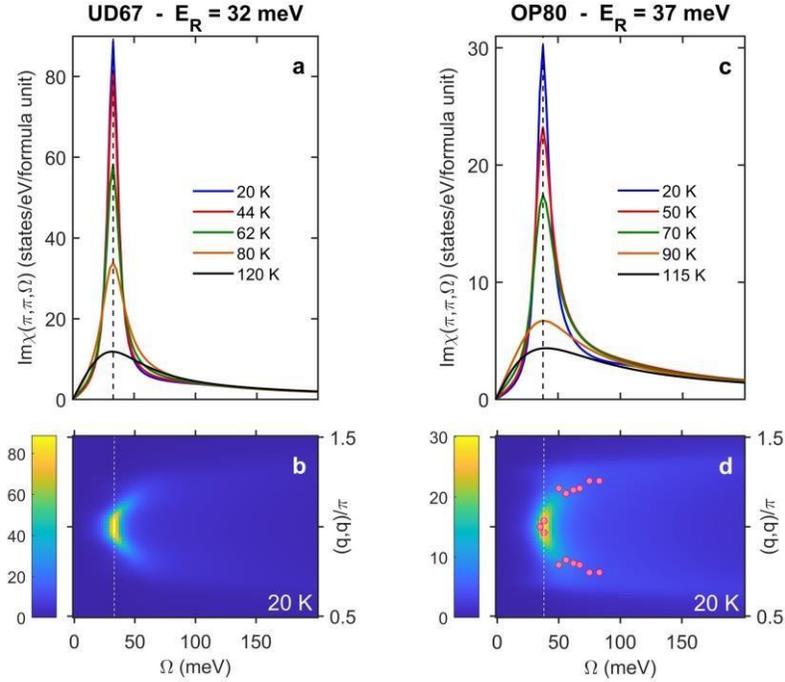

Figure 2: Temperature and doping dependence of the inelastic spin response in Bi2212 calculated from ARPES data. (**a**), (**c**) Calculated spin susceptibility for both thin films (UD67 and OP80) as a function of energy at the commensurate wave vector ($\pi,\pi$) for various temperatures. (**b**), (**d**) Dispersion of magnetic excitations along the diagonal direction at 20 K for UD67 and OP80 samples, respectively. Pink symbols in (**d**) represent INS measurements only above the resonance energy from optimally doped Bi2212 at 10 K (adapted from ref. [25]). Colour bars in (**b**) and (**d**) are in units of (states/eV/formula unit) and dashed lines indicate the resonance energies.

With the experimentally derived values of $U_0$ and the functions $G(\mathbf{k},\omega_n)$ and Im$\chi(\mathbf{q},\omega)$, equation (1) can be solved after obtaining the Matsubara frequency representation for the spin response function:

$$\chi(\mathbf{q},\omega_m) = -\frac{1}{\pi}\int d\Omega \frac{\mathrm{Im}\chi(\mathbf{q},\Omega)}{i\omega_m - \Omega}. \tag{9}$$

For the linearized Bethe-Salpeter equation, we considered only data sets near and above $T_c$, and solved for the leading eigenvalue and eigenvector $\Phi(\mathbf{k},\omega_n,T)$ using the power method [30] (see Supplementary Information). Figs. 3a,b show the momentum dependence of the eigenvector at $\omega_n = 0$ along the direction $(\pi,0) \to (0,\pi)$ for both thin films close to $T_c$. (Results for the OP91 crystal are presented in Supplementary Information.) The eigenvector changes sign and closely follows a cosθ dependence, exhibiting a d-wave character.

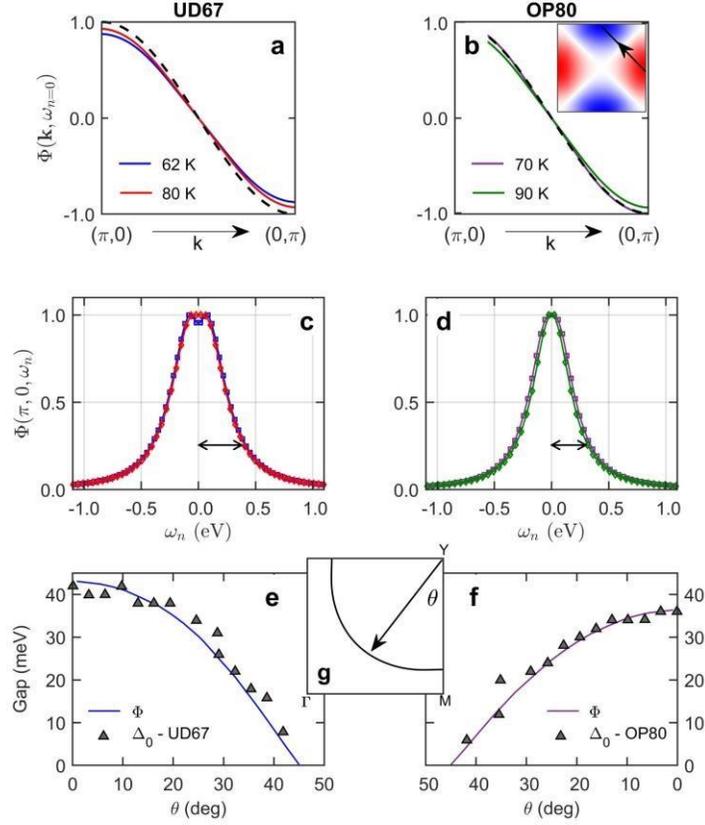

Figure 3: Solution of the Bethe-Salpeter equation and symmetry properties of the order parameter near $T_c$. (**a**) Momentum dependence along the $(\pi,0) \to (0,\pi)$ direction (as shown in the inset of (**b**)) and (**c**) energy dependence of the order parameter $\Phi$ at $(\pi,0)$ for UD67 sample at $T = 62, 80$ K. (**b**), (**d**) same as (**a**),(**c**) but for the OP80 sample at $T = 70, 90$ K. (**e**) Comparison of ARPES gap ($\Delta_0$, gray triangles) for UD67 at 20 K with the corresponding $\Phi(\mathbf{k}, k_B T)$ (at 62 K, solid line) along the Fermi surface shown in (**g**). (**f**) same as (**e**), but the ARPES gap is for the OP80 sample at 20 K and the pairing eigenfunction was calculated from the OP80 data at 70 K. In (**e**), (**f**) the pairing eigenfunctions were re-scaled to match the ARPES gaps at $\theta = 0°$ and the gap values were adapted from ref. [26]. The curve colours are consistent throughout each column of panels. The black, dashed line in (**a**) and (**b**) is the d-wave function $\frac{1}{2}(\cos k_x - \cos k_y)$ along the path $k_y = \pi - k_x$.

For the same data sets, Figs. 3c,d show the (Matsubara) energy dependence of the normalized eigenvector at $(\pi,0)$. The eigenvector is even in frequency, with a characteristic energy scale of the order of 250 meV. Following the Hubbard model calculations of Maier et al. [8], where this characteristic energy is $\sim 2J$, we extract values for the in-plane exchange energy $J \sim 125$ meV, in agreement with estimates obtained from INS for $La_{2-x}Sr_xCuO_4$ and $YBa_2Cu_3O_{6+x}$ [31]. The energy dependence of $\Phi$ implies a retardation in time of the pairing interaction $(2J)^{-1} \sim 10$ fs, as has been observed in ultrafast optical spectroscopy measurements [32]. Finally, the calculated $\Phi$ allows us to recover the superconducting gaps measured in ARPES along the Fermi surface (Fig. 3g), as shown in Fig. 3e (for UD67 at 62 K) and Fig. 3f (for OP80 at 70 K), where the ARPES gap values (gray triangles) were obtained for both films at 20 K [26].

Figure 4 summarizes our results for the leading eigenvalues at temperatures slightly below and above $T_c$, for both thin films and the single crystal. For the three samples, the eigenvalues increase as $T$ is lowered toward $T_c$, as observed in some Hubbard model calculations [7, 8, 33]. This temperature dependence is a direct consequence of using temperature-dependent, experimental spectral functions. Our main result is that, to the accuracy of the present calculation (see Supplementary Information), the eigenvalues are essentially equal to unity near $T_c$.

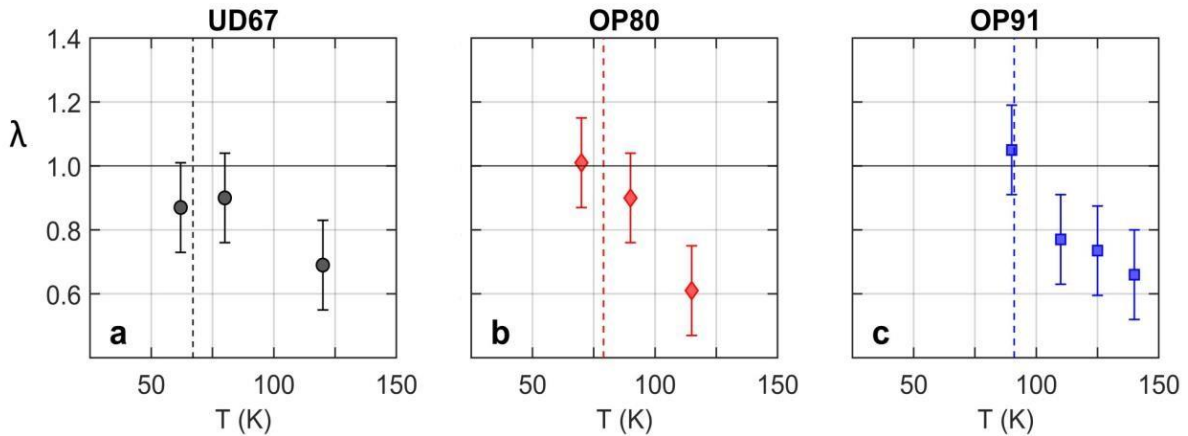

Figure 4: Temperature dependence of leading Bethe-Salpeter eigenvalues for three Bi2212 samples. The dashed lines indicate the critical temperatures for the three samples. The error bars are dominated by the uncertainty in the coupling energy $U_0$, coming from the finite resolution of INS measurements and possible deviations of the $(\pi,\pi)$ peak energy from the value given by equation (8) (see Supplementary Information).

These results show that the magnetically mediated pairing interaction, constructed from ARPES data within the RPA, is sufficiently strong to mediate high-temperature, d-wave superconductivity in the cuprates at optimal and sub-optimal doping. We note that the emergence of superconductivity in the underdoped sample (UD67) arises from the increased strength of the pairing interaction, which compensates for the loss of low-energy electronic states in the pseudogap region. This result is consistent with the notion that AFM correlations promote high-$T_c$ superconductivity in a "dry Fermi sea" [7].

**Methods**

Bi2212 thin films were epitaxially grown by RF magnetron sputtering [34] and measured with 22 eV photons at the U1 undulator beamline at the Synchrotron Radiation Center in Madison, Wisconsin. The spectra were collected with a Scienta R2002 spectrometer, with an energy resolution of 15 meV. The films were mounted with the Cu-O bond parallel to the photon polarization and cleaved in situ at a pressure below $2 \times 10^{-11}$ Torr. The optimally doped single crystal was grown by the traveling solvent, floating zone technique, and measurements were performed at the University of Illinois at Chicago, using the unpolarized He-I line (21.2 eV) from a Helium discharge lamp. Photoelectrons were analyzed with a Scienta R4000 spectrometer, the energy resolution of the system was set at 25 meV, and the sample was measured at a pressure below our measurement limit of $2 \times 10^{-11}$ Torr.

**Data availability** The authors declare that the main data supporting the findings of this study are available from the corresponding authors upon reasonable request.

**Code availability** The codes employed in this study are available from the authors upon reasonable request.

**Acknowledgements** U.C. acknowledges support from the National Science Foundation under grant number DMR-1454304. D.K.M. acknowledges support from the U. S. Department of Energy, Office of Science, Basic Energy Sciences, under Award No. DE-FG02-05ER46225.

**Author contributions** F.R, U.C., J.C.C., and D.K.M. conceived the project, carried out the work, and wrote the manuscript. J.C.C. and U.C. provided the thin film data, and F.R. and H. X. collected the single crystal data. G. D. G. provided the Bi2212 single crystal.

**Competing financial interests** The authors declare no competing financial interests.

# Supplementary Information for *Coupled electronic and magnetic excitations in the cuprates and their role in the superconducting transition*

F. Restrepo[1], U. Chatterjee[2], G. D. Gu[3], H. Xu[1], D. K. Morr[1,4], and J. C. Campuzano[1]

[1] *Department of Physics, University of Illinois at Chicago, Chicago, Illinois 60607, USA*

[2] *Department of Physics, University of Virginia, Charlottesville, Virginia 22904, USA*

[3] *Condensed Matter Physics and Materials Science Division, Brookhaven National Laboratory, Upton, New York 11973, USA*

[4] *James Franck Institute, University of Chicago, Chicago, Illinois 60637, USA*

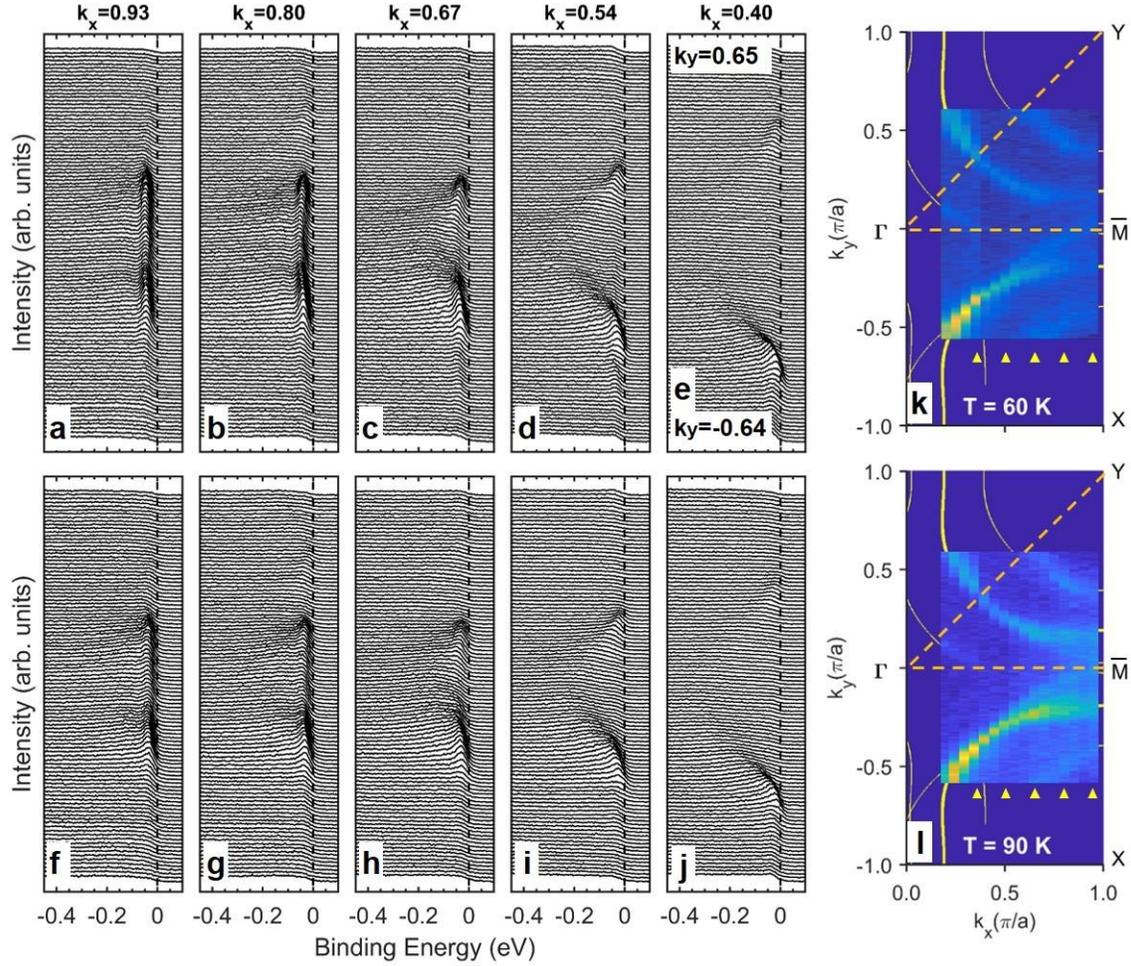

**Supplementary Figure 1: Selection of ARPES data from the Bi2212 single crystal below and near $T_c$.** **(a)** EDCs located at coordinates **k** = (0.93,−0.64 → 0.65)$\pi/a$ with $T$ = 60 K (slice indicated by rightmost triangle in panel k). EDCs in panels **(b)** - **(e)** correspond to successively lower values of $k_x$ (indicated by triangles in panel **(k)**). Panels **(f)** - **(j)** show the same EDCs but at $T$ = 90 K. **(k)** and **(l)** are the Fermi surfaces measured at $T$ = 60 K and $T$ = 90 K, respectively, laid over a tight binding fit to the dispersion [1].

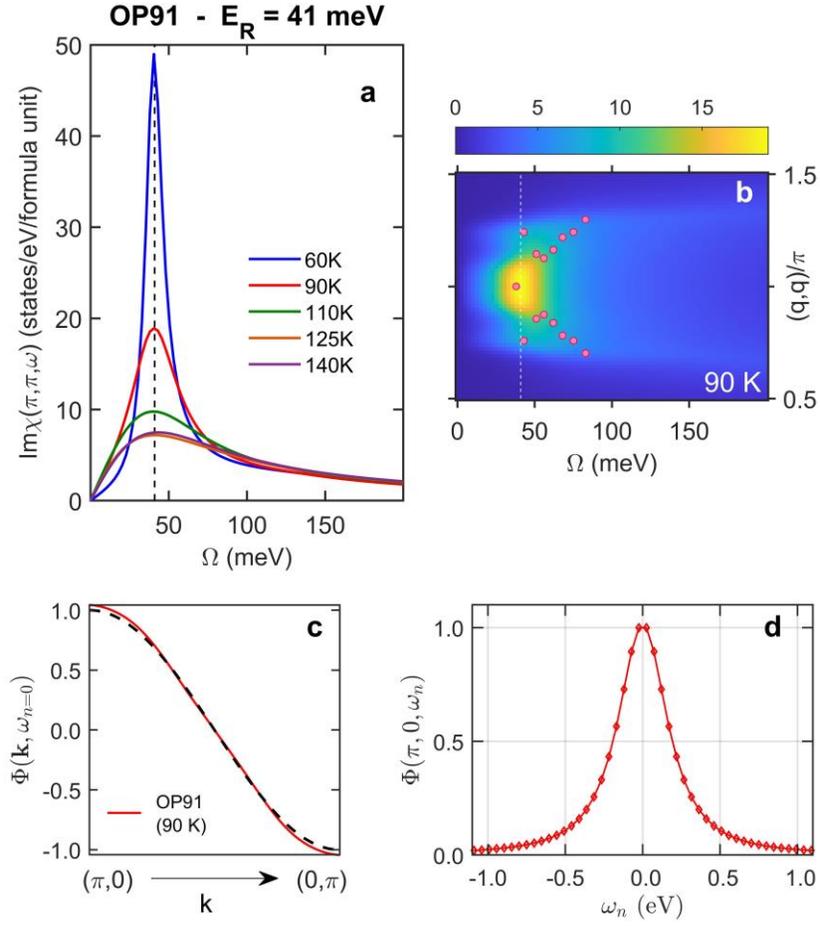

**Supplementary Figure 2: Additional calculations for the OP91 Bi2212 single crystal.** **(a)** ARPES-derived spin susceptibility at the commensurate wave-vector $(\pi,\pi)$ for temperatures between 60 K (below $T_c$) and 140 K ($\sim T^*$). **(b)** Energy-momentum dispersion of the spin susceptibility along the $(\pi/2,\pi/2) - (3\pi/2,3\pi/2)$ direction. Pink symbols represent INS measurements (only above the resonance energy) from optimally doped Bi2212 at 100 K (adapted from ref. [2]). **(c)** Generalized order parameter for 90 K data along $(\pi,0) - (0,\pi)$ direction (dashed line is the pure d-wave function $\cos k_x - \cos k_y$). **(d)** Matsubara energy dependence of order parameter at $(\pi,0)$ at 90 K.

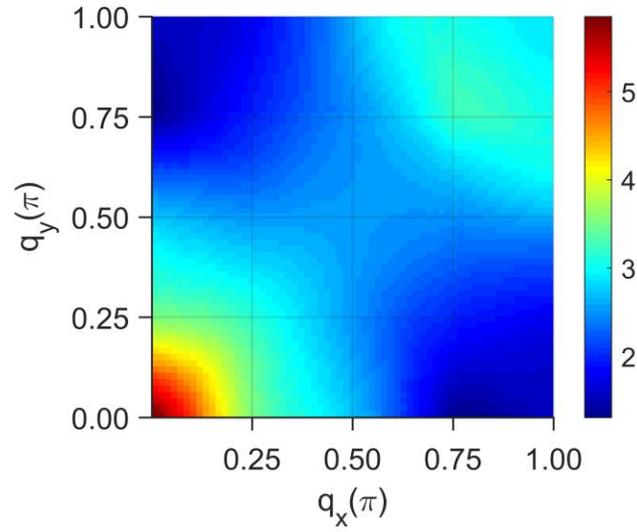

**Supplementary Figure 3: (Color online) Real part of the bare susceptibility at $\omega = 0$ for an optimally doped Bi2212 sample at 140K (normal state)**. The color bar is in units of states/eV/formula unit.

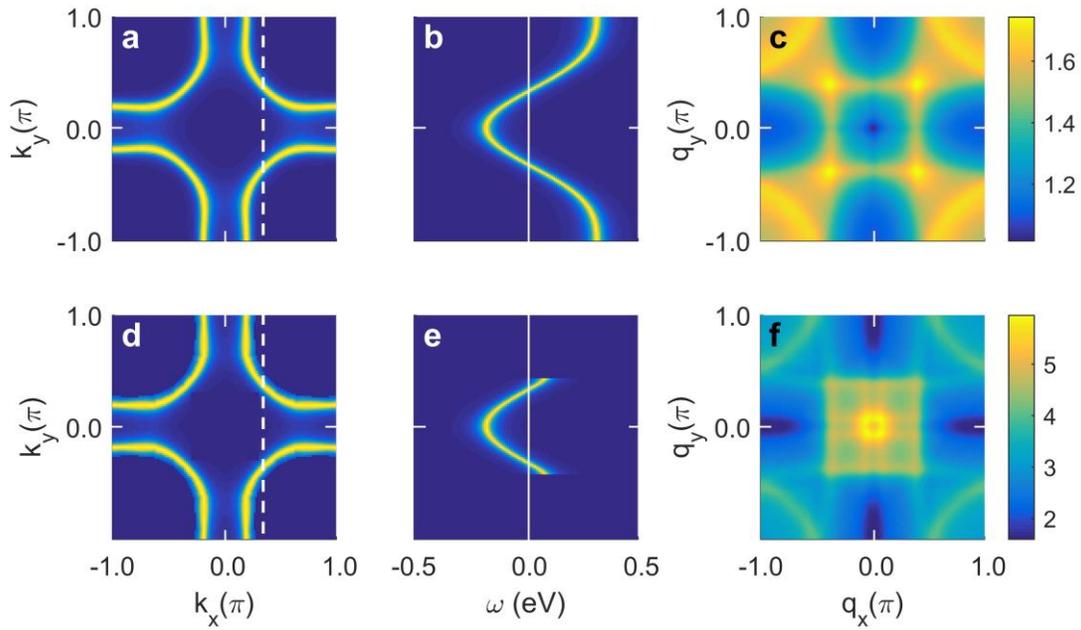

**Supplementary Figure 4: (Color online) Effects of symmetrization on the bare susceptibility $\chi_0$. (a)** FS of the true spectral function. **(b)** Dispersion along direction

indicated by white vertical line in **(a)**. **(c)** Calculated $\chi_0(\mathbf{q},\omega = 0)$. **(d)-(f)** Same as **(a)-(c)** but the spectral function has been symmetrized "by hand" and some patches of the BZ have been removed. Color bars in **(c)** and **(f)** are in units of states/eV/formula unit.

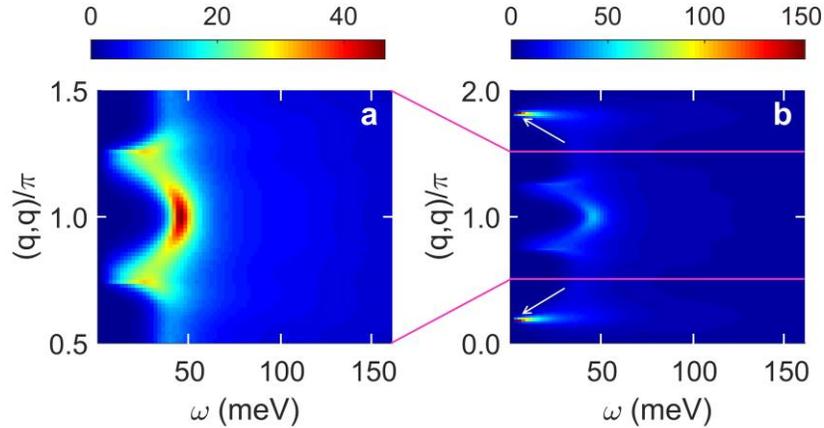

**Supplementary Figure 5: (Color online) Symmetrization artifacts in the interacting spin susceptibility of near-optimally doped Bi2212. (a)** ARPES-derived magnetic excitation spectrum along diagonal direction from $(0.5\pi, 0.5\pi)$ to $(1.5\pi, 1.5\pi)$ **(b)**. Same as **(a)**, but in a momentum range $(0,0)-(2\pi,2\pi)$. Note the difference in color bar scales (states/eV/formula unit).

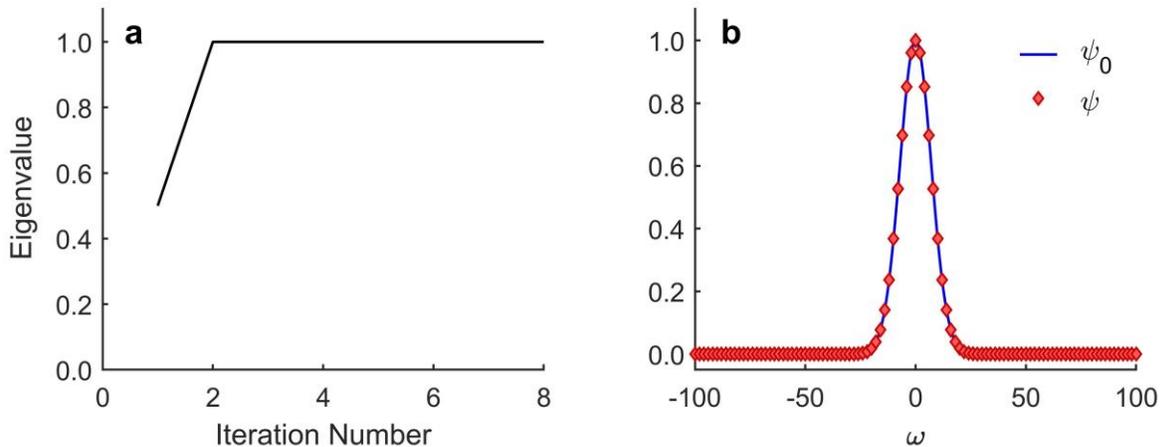

**Supplementary Figure 6: (Color online) (a)** Behavior of the eigenvalue as a function of iteration number in the power method for the sinc-convolution problem. **(b)** Initial eigenfunction guess (red diamonds) and calculated eigenfunction (blue line).

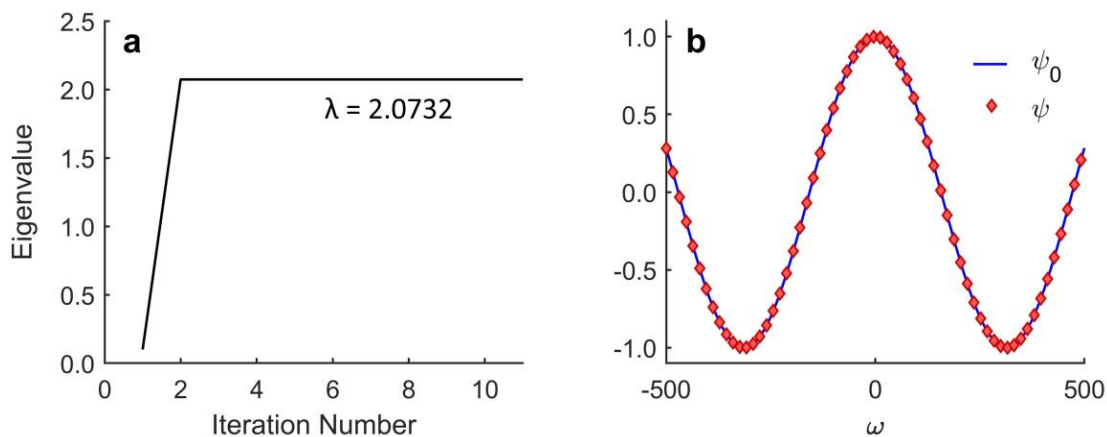

**Supplementary Figure 7: (Color online) (a)** Behavior of the eigenvalue as a function of iteration number in the power method. **(b)** Initial eigenfunction guess (blue line) and calculated eigenfunction (red diamonds).

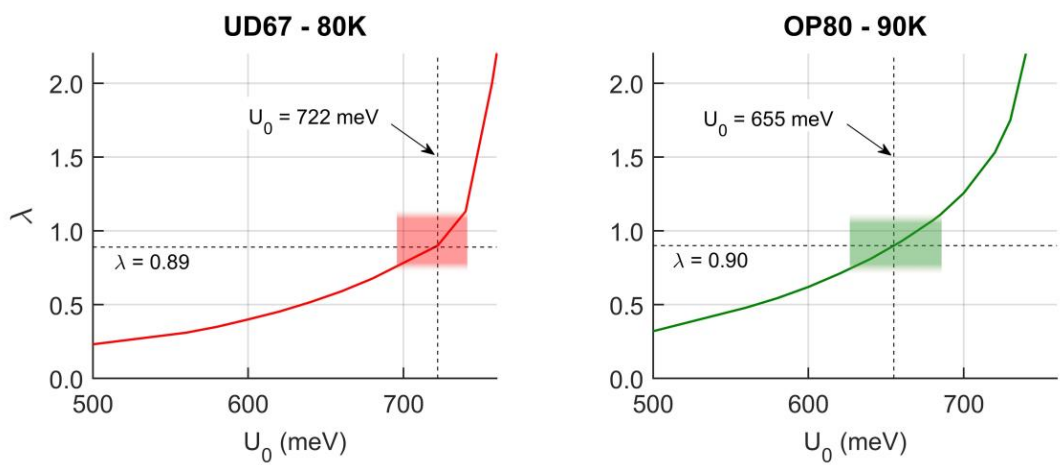

**Supplementary Figure 8: (Color online) Leading eigenvalue as a function of coupling energy for both Bi2212 thin films above (but close to)** $T_c$**.** Shaded rectangles indicate uncertainties in $U_0$ and $\lambda$.

**Supplementary Table 1: Comparison between the coupling energies ($U_0$ and $U$) and between the corresponding eigenvalues ($\lambda_0$ and $\lambda$ respectively) near $T_c$**

| Sample | T (K) | $U_0$ (meV) | $U$ (meV) | $\lambda_0$ | $\lambda$ |
|---|---|---|---|---|---|
| UD67 | 80 | 720 | 775 | 0.89 | **1.02** |
|  | 62 | 741 | 820 | 0.87 | **1.1** |
| OP80 | 90 | 655 | 635 | 0.9 | **0.86** |
|  | 70 | 660 | 635 | 1.01 | **0.93** |
| OP91 | 90 | 585 | 525 | 1.05 | **0.84** |

# 1   Additional data and results for the OP91 single crystal

In the main text we showed a small sample of the data sets employed in this work for the calculation of the pairing interaction and the solution of the Bethe-Salpeter Equation (BSE). Supplementary Figure 1 shows a more detailed example of the kind of data used in said calculations, collected from the OP91 single crystal at 60 K and 90 K. Panel (*k*) shows the momentum dependence of electronic excitations at the Fermi energy below $T_c$ for half the Brillouin Zone (BZ). As in the UD67 sample data (shown in the main text), the main and superlattice bands are visible, but the latter are more intense due to the unpolarized UV radiation used for this sample. The dashed triangle indicates the portion of momentum space selected for reflection about the symmetry axes of the zone to obtain the spectra throughout the BZ. Panels (*a*)–(*e*) show stacked Energy Distribution Curves (EDCs) along selected vertical slices of momentum space, starting near the ($\pi,k_y$) line and ending in a slice containing the node (see yellow triangles in panel (*k*)). For both temperatures, the dispersion of the antibonding band is evident from the EDCs, as is the closing of the superconducting gap near the node. Even though the EDCs are shown for binding energies up to 450 meV, in the calculations in the main text we used the entire EDCs, sampled up to 520 meV for the OP91 sample.

Using these data, we followed the same procedure as in the main text to calculate the spin susceptibility and solve the BSE. Supplementary Figure 2 summarizes the results for the OP91 crystal, emphasizing those obtained at 90 K ($\sim T_c$). In real frequencies, Im$\chi$ again broadens and loses intensity as the temperature is increased (Supplementary Figure 2a), while the upward dispersion branch is still visible in Supplementary Figure 2b. Pink symbols in this panel represent INS data from optimally doped Bi2212 at 100 K [2]. The pairing eigenfunction Φ also changes sign upon crossing the BZ diagonal and decays with a similar energy scale as in the thin films ($\sim 250$ meV), as shown in Supplementary Figure 2c, d.

## 2 The anomalous Green's function

In the superconducting state, the particle-hole bubble $\chi_0$ should include a term quadratic in the anomalous Green's function, which we call *FF*. We ignored this term in this work for three reasons:

> As pointed out by Chatterjee et al. [3], in the superconducting state, the calculated dynamic susceptibilities vary only slightly when one includes the anomalous Green's function. Inclusion of the *FF* term will certainly influence the *bare* susceptibility, but an appropriate readjustment of the coupling constant $U_0$ will keep the resonance in the interacting susceptibility at the correct energy.

> While we considered several data sets below $T_c$ to illustrate the performance of the RPA in constructing the spin fluctuation propagator, for the calculation of the BetheSalpeter eigenvalue we restricted our analysis to temperatures slightly below and above $T_c$, where the *FF* term is not important.

> Similar results for the eigenvalues were obtained from an independent method of obtaining the prefactor $U^2$ in the pairing interaction $V$, as shown in Supplementary Section 4.

However, it is instructive to examine the effect of this term on the RPA energies $U_0$ in the superconducting state. To this end, we first expressed the real part of the bare susceptibility as $\text{Re}(\chi_0) = \text{Re}(\chi_0^G) + \text{Re}(\chi_0^F)$ (where the second term is the anomalous contribution coming from the *FF* term) and defined the ratio $\alpha = \text{Re}(\chi_0^F/\chi_0^G)$. Since $\chi_0^F$ is unknown, so is $\alpha$. However, in a d-wave superconductor, $\text{Re}(\chi_0^F)$ and $\text{Re}(\chi_0^G)$ add constructively at the commensurate wave-vector, so we expect this ratio to be positive at $(\pi,\pi)$. From the pole condition for the resonance, $1 - U_0\text{Re}(\chi_0) = 0$, we see that an enhancement of the real part of the bare susceptibility reduces $U_0$. To estimate this reduction, we computed $\text{Re}\chi_0^G$ as in the main text (using ARPES data below $T_c$) and expressed the real part of the *total* bare susceptibility as

$$\text{Re}(\chi_0) = \text{Re}(\chi_0^G)(1 + \alpha_{\text{BCS}}), \tag{1}$$

where the ratio $\alpha$ was calculated using BCS normal and anomalous spectral functions [4] and a tight-binding fit for the Bi2212 dispersion [1]. We considered superconducting state spectral functions with a Lorentzian lineshape, sampled over a momentum grid of 128×128 points and energies ranging from −1.5 to 1.5 eV. We also allowed for a finite width ($\Gamma$) of the spectral function and studied the effect of varying this width. The convergence factor $\delta$ was set to 10 meV, and superconductivity was included through the d-wave gap

$$\Delta(\mathbf{k}) = \frac{\Delta_0}{2}(\cos k_x - \cos k_y), \tag{2}$$

with $\Delta_0$ = 35 meV. Since we were only interested in the effect on the coupling constant $U_0$, we compared results only at the commensurate wave-vector $(\pi,\pi)$. For $\Gamma$ = 40 meV, $\alpha$ = 0.05, leading to a 5% decrease in $U_0$. For $\Gamma$ = 20 meV, $\alpha$ = 0.10, leading to a 10% decrease in $U_0$. Finally, for $\Gamma$ = 0 meV, $\alpha$ = 0.2, leading to a 20% decrease in $U_0$. The values of $U_0$ reported in the main text do not take this reduction (∼ 10% for the more physically relevant case) into account because the data employed were collected near $T_c$.

# 3 Obtaining the spin susceptibility from ARPES data: the choice of RPA coupling $U_q$

Supplementary Figure 3 shows the ARPES-derived *bare* susceptibility $\chi_0(\mathbf{q},\omega = 0)$ for an optimally doped Bi2212 sample in the normal state[5]. The results agree qualitatively with those in ref. [1]. There is a pronounced, square-like signal enclosing the $\mathbf{q} = 0$ point, as well as an incommensurate response around $\mathbf{Q} = (\pi,\pi)$. However, there is also significant response at $\mathbf{q} = 0$, which is not seen in model calculations (Figure 1 of ref. [1]). This response was found to be an artifact of the Fermi cutoff on the spectral function.

To demonstrate the fictitious origin of the $\mathbf{q} = 0$ peak, we calculated $\chi_0$ using Lorentzian spectral functions with tight binding fits for the dispersion of Bi2212. The procedure was essentially the same as in Supplementary Section 2, but setting $\Delta_0 = 0$ (for the normal state) and $\Gamma = 20$ meV. Supplementary Figures 4a-b show the Fermi Surface and the dispersion for the true spectral function (without symmetrization). The antibonding band disperses (along the dashed line in panel **a**) all the way to the BZ boundary. The bare susceptibility (at $\omega = 0$) for this spectral function (Supplementary Figure 4c) closely resembles similar calculations presented in ref. [1]. Note the absence of spectral weight at $\mathbf{q} = (0,0)$. The second row shows similar calculations but now the spectral function has been multiplied by the Fermi function and symmetrized, to simulate the analysis we did with experimental data. The dispersion does not cover the entire BZ because particle-hole symmetry is only assumed to hold for small binding energies (in this case, we took 80 meV about the Fermi energy). We see that the resulting bare susceptibility (panel f) has significant spectral weight at the zone center (note also the difference in color bar scales). The artifact is thus a direct consequence of not having access to the complete spectral function (including the unoccupied states), a shortcoming which can only partially be mitigated through symmetrization.

This artifact is also apparent in the *interacting* susceptibility, approximated via RPA as [7]:

---

[5] The results in this section were derived from ARPES data (from a near-optimally doped Bi2212 single crystal) presented in earlier publications, e.g. refs. [5, 6], not the data used in the main text. We use these data here only for illustration purposes.

$$\chi(\mathbf{q},\omega) = \frac{\chi_0(\mathbf{q},\omega)}{1 - U_q\chi_0(\mathbf{q},\omega)}. \tag{3}$$

If $U_q$ is momentum-independent, $\chi(\mathbf{q},\omega)$ possesses an anomalously strong peak near $\mathbf{q} = 0$, which is not seen in INS measurements. In Supplementary Figure 5 we compare the magnetic dispersion for Im$\chi$ calculated with a momentum-independent coupling $U_q = U_0$ from $\mathbf{q} = (\pi/2,\pi/2)$ to $\mathbf{q} = (3\pi/2,3\pi/2)$ (as is frequently reported in the literature) with the same quantity plotted from $\mathbf{q} = (0,0)$ to $\mathbf{q} = (2\pi,2\pi)$. The ARPES data employed correspond to an optimally doped Bi2212 sample in the superconducting state[6]. In the restricted momentum view, $U_0$ yields a spin response with rich momentum structure, resembling the famous "hourglass" shape reported in the literature on other cuprates. However, the wider momentum view reveals two peaks at low energy near $\mathbf{q} = (0,0)$ and $\mathbf{q} = (2\pi,2\pi)$ three times larger than the rest of the signal (indicated by white arrows in Supplementary Figure 5b). The spin response thus constructed displays an inordinate amount of spectral weight at momenta where experiments do not detect it. This artifact was removed by the choice of coupling function $U_q$ defined in the main text, which suppresses the response around $\mathbf{q} = n(2\pi,2\pi)$. We emphasize that such a choice of $U_q$ was not made to favor spin-fluctuation-induced superconductivity models, but was rather motivated by the fact that it rectifies the small-$\mathbf{q}$ intensity problem and yields a spin susceptibility in better agreement with experimental observations.

## 4 Alternative estimate of the spin-fermion coupling energy

It has been pointed in ref. [6] that the the effective electron-electron interaction should be written as

$$V(\mathbf{q},\omega) = \frac{3}{2}U^2\frac{\chi_0(\mathbf{q},\omega)}{1 - U_q\chi_0(\mathbf{q},\omega)}, \tag{4}$$

---

[6] These data were collected from the same sample as those in Supplementary Figure 3, but in the superconducting state.

where the coupling energy $U$ may differ from $U_0$ due to vertex corrections. Following ref. [6], here we show how this energy can be obtained from the electron self-energy. In this section, we only report results in the important region near $T_c$.

Assuming the single-particle renormalizations are due to $V$, we can estimate the self-energy via

$$\mathrm{Im}\Sigma(\mathbf{k},\omega) = \frac{1}{N}\Sigma_{\mathbf{q}}\int d\Omega [n(\Omega) + f(\Omega - \omega)] \mathrm{Im}V(\mathbf{q},\Omega) A(\mathbf{k}-\mathbf{q},\omega-\Omega), \quad (5)$$

where $n(\omega)$ and $f(\omega)$ are the Bose and Fermi distributions, respectively, and the integral is cut off at $\Omega = 0.4$ eV. By Kramers-Kronig relations, we can then obtain the real part of the self-energy and the nodal Fermi velocity renormalization,

$$Z = \frac{v_F^0}{v_F} = 1 - \frac{d}{d\omega}Re\Sigma(k_n,\omega)|_{\omega=0}, \quad (6)$$

where $k_n$ is the nodal Fermi momentum and $v_F^0$ and $v_F$ are the bare and interacting nodal Fermi velocities, respectively. Based on the $T$-linear behavior of $v_F$ inferred from laser ARPES data from optimally doped Bi2212 [8], we estimate that, for the OP80 sample, $Z(70K) \sim 2.1$ and $Z(90K) \sim 2$, with similar values for the single crystal OP91. For the UD67 sample we have $Z(62K) \sim 2.4$ and $Z(80K) \sim 2.3$ [9]. Using the measured spectral functions and the spin susceptibilities calculated in the main text, we computed $Z$ while adjusting $U$ until we obtained the values above. For all samples, $U$ and $U_0$ differ by at most 11%, which is somewhat lower than the deviation reported in [6]. Using these coupling energies, the eigenvalues are still essentially equal to unity near $T_c$, as shown in Supplementary Table 1. For comparison, we also included the eigenvalues obtained by assuming $U = U_0$, as in the main text (labeled $\lambda_0$).

# 5 The power method for the Bethe-Salpeter Equation

The method employed in solving the eigenvalue problem

$$\lambda(T)\Phi(T) = \hat{O}(T)\Phi(T) \tag{7}$$

with

$$\hat{O}(T)\Phi(\mathbf{k},\omega_n,T) = -\frac{T}{N}\sum_{\mathbf{k}',\Omega_n} V(\mathbf{k}-\mathbf{k}',i\omega_n-i\Omega_n,T)G(\mathbf{k}',i\Omega_n,T)G(-\mathbf{k}',-i\Omega_n,T)\Phi(\mathbf{k}',\Omega_n,T) \tag{8}$$

followed very closely that of Monthoux [10]. In assessing whether a hypothetical pairing interaction is strong enough to mediate superconductivity at a given temperature, it is sufficient to focus on the largest eigenvalue of $\hat{O}$. To this end, we used the *power method*, which is an iterative procedure to calculate the maximum eigenvalue of an operator. There are several equivalent ways of carrying out this procedure; in this work, the implementation goes as follows:

> We start with a guess for the eigenfunction $\Phi$, which is a generalized gap function reflecting the momentum and energy dependence of the pairing interaction. Motivated by the well-established d-wave form of this interaction, we take $\Phi_0(\mathbf{k},\omega_n) = \cos(k_x) - \cos(k_y)$ as our starting guess.

> With this guess, we perform the first iteration $F_1 = \hat{O}\Phi_0$ and factor out the largest entry in $F_1$, so that $F_1 = \lambda_1\Phi_1$, where $\lambda_1 = \max(F_1)$ and $\Phi_1 = F_1/\lambda_1$.

> We then use $\Phi_1$ as input to the second iteration $F_2 = \hat{O}\Phi_1$ to obtain $\lambda_2 = \max(F_2)$ and $\Phi_2 = F_2/\lambda_2$, as before.

> The above process is repeated until the iteration ($N$) is reached where the difference $\lambda_N - \lambda_{N-1}$ is negligible. At this point, the function $\Phi_N$ is the normalized eigenfunction of $\hat{O}$ corresponding to the largest eigenvalue $\lambda_N$.

Since Supplementary Equation (8) is essentially a convolution equation, the operations $\hat{O}\Phi_j$ were performed using Fourier techniques for increased computational speed. The numerical efficiency afforded by the convolution structure of the BSE was our main reason for studying the pairing problem in Matsubara frequencies.

From the Lehmann representation of the Green's function, it can be seen that $P(k,\omega_n) = G(k,\omega_n)G(-k,-\omega_n)$ is even in Matsubara frequency, real, and positive. For the Matsubara spin response, the property $\text{Im}\chi(q,-\Omega) = -\text{Im}\chi(q,\Omega)$ implies that $\chi(q,\omega_m)$ is also even in $\omega_m$, real and positive. These properties are not only useful as consistency checks of the numerical procedure, but also enable us to use the power method, which breaks down when the eigenvalues are complex.

We now present two examples of convolution equations which are one-dimensional analogues of the BSE considered in this work. We begin with an equation frequently found in signal processing applications:

$$\int_{-\infty}^{\infty} \text{sinc}(\omega - \omega')\psi(\omega')d\omega' = \lambda\psi(\omega). \tag{9}$$

The convolution in Supplementary Equation (9) can be expressed as

$$F^{-1}\{F[\text{sinc}(\omega)]F[\psi(\omega)]\}, \tag{10}$$

where $F$ and $F^{-1}$ denote Fourier and inverse Fourier transforms, respectively. Since $F[\text{sinc}(\omega)]$ is just a "boxcar" function, if $F[\psi(\omega)]$ is narrower than this function, then $F[\text{sinc}(\omega)]F[\psi(\omega)] \sim F[\psi(\omega)]$. This situation corresponds to a band pass filter which is broader than the input signal, in which case the signal is passed unaffected. Therefore, we expect $\psi(x)$ to be an eigenfunction of the sinc-convolution operator with $\lambda = 1$. To test this, we choose a Gaussian test function $\psi_0(x) = e^{-0.1x^2}$ for our initial guess and set $\lambda_0 = 0.5$ initially. Because in this simple case the initial guess *is* the eigenfunction, the power method immediately converges to the eigenfunction $\psi = \psi_0$ with $\lambda = 1$, as shown in Supplementary Figure 6.

As a second example, consider another convolution product $f(\omega)*\psi(\omega)$, where $\psi(\omega) = e^{\alpha\omega}$ ($\alpha$ is a complex number). It can be shown that

$$\int_{-\infty}^{\infty} f(\omega')\psi(\omega - \omega')d\omega' = \lambda_\alpha \psi(\omega), \tag{11}$$

with

$$\lambda_\alpha = \int_{-\infty}^{\infty} f(\omega) e^{-\alpha\omega} d\omega \qquad (12)$$

Thus, $\psi(\omega)$ is the eigenfunction of a convolution operator (with $f(\omega)$ as a kernel) with an eigenvalue given by Supplementary Equation (12). To test our procedure, we take a gaussian "test" kernel $f(x) = e^{-\omega^2}$, and assume $\alpha = 0.01i$ (this small value was chosen so that the imaginary components in Supplementary Equation (11) vanish). The eigenvalue obtained from Supplementary Equation (12) is 2.0732. To apply the power method, we take $\psi_0(\omega) = e^{(0.01i)\omega}$ and $\lambda_0 = 0.1$ as our starting guesses and compute the convolution integral with Fourier transforms (as done with the BSE in this work). Supplementary Figure 7a shows the eigenvalue as a function of iteration number. Again, we see that $\lambda$ converges very quickly to the expected 2.0732. Moreover, Supplementary Figure 7b shows that the calculated eigenfunction again matches the initial guess $\psi_0(\omega)$.

## 6  Uncertainties in the eigenvalues

The leading eigenvalue in the BSE depends on the spin-fermion coupling through the factor $U_0^2$ and the denominator of the RPA expression for the spin susceptibility. This coupling, in turn, was fixed from the empirical relation

$$E_R = 5.4 k_B T_c. \qquad (13)$$

Therefore, we estimate the uncertainty in the eigenvalue by relaxing the above relation and introducing the uncertainty $\Delta E_R = 10$ meV. This rather large variation was chosen to account not only for the INS energy resolution (< 5 meV), but also for the fact that the energy of the $(\pi,\pi)$ peak may vary somewhat with temperature (as seen, for example, in ref. [11]). Supplementary Figure 8 shows plots of the leading eigenvalue as a function of the coupling $U_0$ for both thin films above (but close to) $T_c$. The dashed vertical lines represent the values of $U_0$ used in the main text and for which the $(\pi,\pi)$ peak energy was exactly given by Supplementary Equation 13. The horizontal dashed lines are the resulting eigenvalues (reported in the main text). The horizontal dimension of the shaded rectangles represents the range of $U_0$ values for which the peak energy moves from $E_R + \Delta E_R$ meV (lowest $U_0$) to $E_R$ $-\Delta E_R$ meV (highest $U_0$). The vertical dimension represents the resulting uncertainty in the eigenvalue. Thus, we estimate a 15%–20% variation in the leading eigenvalues reported in the main text.